\documentclass{article}

\usepackage{PRIMEarxiv}

\usepackage[utf8]{inputenc} 
\usepackage[T1]{fontenc}    
\usepackage{hyperref}       
\usepackage{url}            
\usepackage{booktabs}       
\usepackage{amsfonts}       
\usepackage{nicefrac}       
\usepackage{microtype}      
\usepackage{lipsum}
\usepackage{graphicx}
\graphicspath{{media/}}     

\title{Inferential Moments of Uncertain Multivariable Systems
}

\author{
 Kevin Vanslette \\
  Raytheon BBN \\
  10 Moulton St., Cambridge, MA 02138, United States\\
  \texttt{kevin.vanslette@rtx.com} \\
}

\begin{document}
\maketitle

\begin{abstract}
This article expands the framework of Bayesian inference and provides direct probabilistic methods for approaching inference tasks that are typically handled with information theory. We treat Bayesian probability updating as a random process and uncover intrinsic quantitative features of joint probability distributions called inferential moments. Inferential moments quantify shape information about how a prior distribution is expected to update in response to yet to be obtained information. Further, we quantify the unique probability distribution whose statistical moments are the inferential moments in question. We find a power series expansion of the mutual information in terms of inferential moments, which implies a connection between inferential theoretic logic and elements of information theory. Of particular interest is the inferential deviation, which is the expected variation of the probability of one variable in response to an inferential update of another. We explore two applications that analyze the inferential deviations of a Bayesian network to improve decision-making. We implement simple greedy algorithms for exploring sensor tasking using inferential deviations that generally outperform similar greedy mutual information algorithms in terms of root mean squared error between epistemic probability estimates and the ground truth probabilities they are estimating.
\end{abstract}


\section{Introduction}

Bayesian inference frameworks use Bayesian probability theory for knowledge representation, updating, and reasoning in the presence of uncertainty \cite{Jaynesbook,Catichabook, Cox}; however, there is a sense in which they are incomplete. Rather than working solely within a Bayesian inference framework, practitioners are often forced to draw inferences from information theory to accomplish certain tasks. This is especially prevalent for tasks having to do with optimal sensor tasking \cite{sensortasking, erwin2010dynamic, WilliamsSpencer, jaunzemis2018sensor, yan2020sensor, Adurthi, gerber2022consolidated}, adaptive testing \cite{almond1999graphical,plajner2016student,plajnerquestion}, feature selection \cite{shah2016review,bennasar2015feature,peng2005feature,liu2020exploring, fish2021entropic},  network learning and inference \cite{granger1969investigating,schreiber2000measuring,sun2014causation,sun2015causal,fish2022interaction}, and Bayesian network structure learning \cite{cheng1997learning,kitson2023survey} where inferences are needed to be made about the nature and structure of the probability distributions themselves. In such cases, information theoretic quantities like KL-divergence, entropy, Fisher information, information gain, and most prominently, the mutual information (MI),
\begin{eqnarray}
{MI}[A,B] &=& \sum_{a,b}p(a,b)\ln\Big(\frac{p(a,b)}{p(a)p(b)}\Big),\label{MI}
\end{eqnarray}
are used to make statements about probability distributions as a whole and provide a ranking mechanism in terms of the amount of available information or the amount of mutual dependence between variable sets to assist the inference process \cite{shannon1948mathematical,Cover}. Most information theoretic quantities involve expectation values over the log of probabilities and have been applied widely for reasoning, inference, and modeling uncertain systems in physics, chemistry, biology, engineering, complex systems, inquiry (experimental design, sensor tasking etc.), machine learning, artificial intelligence (AI), autonomy, and economics. These information theoretic tools have the attractive property that they factor when variables are independent.

Many probabilistic inference tasks, however, are solvable within the Bayesian inference framework. Bayesian model learning, calibration, validation, and testing are Bayesian inference procedures that involve fitting parametric models (typically low dimensional physics-based models) to given data in a robust manner by learning model parameter uncertainties and selecting which of a set of candidate predictive models is most probable \cite{gelfand1994bayesian,sivia2006data, geweke2007bayesian, gai2018bayesian, vanslette2020general,tohme2020generalized}. These methods utilize uncertainty propagation (marginalization) and Bayes' Rule for representing solutions to the model ``inverse" problem -- the solutions of which often needed to be approximated via Markov Chain Monte Carlo-like sampling methods. Analogous to Bayesian model testing is Bayesian hypothesis testing and evidence weighting where the probability a hypothesis is true, given evidence, is computed. In the domain of artificial intelligence and machine learning, non-parametric parameter optimization is dominated by negative log-likelihood optimization, which is used to create uncertainty-aware (uncertainty predicting) AI models that ``know when they don't know" \cite{ovadia2019can, lakshminarayanan2017simple, sensoy2018evidential}. Bayesian networks \cite{Pearl,Huang,jensen2007bayesian} offer graphical representations of joint probability distributions and insights into causation via d-separation that are widely used across scientific and engineering domains for making inferences in the presence of uncertainty. Once a probabilistic system has been posited, sensitivity analysis \cite{castillo1997sensitivity,razavi2016new} can be used to explore how small errors in model parameters, evidence, or parameter probabilities in a Bayesian network affect the predicted outputs and downstream decisions drawn from the model. Further, subjective logic \cite{subjectivelogic} offers a method to explore uncertainties of probability distributions from Dirichlet distribution perspective, either through the subjectivity of priors or from a lack of an infinite supply of data. Marginalization, Bayes' rule, and Bayes' theorem are utilized ubiquitously to complete these Bayesian inference tasks.

%
%
%
%

In this article we will investigate the use of Bayesian inference on the process of Bayesian probability updating itself (i.e. Bayes' rule) through the quantification of \emph{inferential moments} and the \emph{inferential probability distribution}. This offers a competitive Bayesian inference-based approach to solve tasks typically handled with information theory. We quantify the sense in which a marginal distribution is the first order inferential moment and a mean estimate of a conditional distribution having unknown conditions. Analogously, the second central inferential moment, the inferential variance, describes the expected variation of a conditional distribution when the conditions themselves are uncertain. We relate the inferential variance to the mean square error between the prior distribution estimate and the true, yet unknown, underlying possible posterior distributions that could arise out of the Bayesian updating process. We construct the inferential probability distribution, which is the unique probability distribution over possible posterior probability updates for a given joint probability distribution that’s moments are the inferential moments in question. In the special case in which inferential updates are data-like, i.e. completely certain measurements, we find that the inferential probability distribution is Bernoulli distributed. We find that inferential moments may be playing a supporting role in some aspects of information theory as the mutual information may be understood as a power series expansion of inferential moments for distributions with loosely correlated variables. This, along with the inferential variance relationship to mean squared error, suggests inferential theoretic logic may be useful for performing inference tasks typically handled with information theory. Rather than ranking which variable is most \emph{informative} from an information theory perspective, we enable the ranking of which variable measurement (or utilization) is expected to result in the best posterior estimate of distribution in a mean square error sense, which arguably is more valuable for estimation and downstream decision making. Further, because this approach is framed in terms of probabilities and their propensity for change, these quantities are understood directly in terms of amounts of probability change rather than amounts of information or numbers of bits, which are typically only understood by experts or in relative terms of one situation having more information than another. 
We explore the application of inferential moments and the inferential probability distribution for improved state (or situation) monitoring as well as to a toy sensor tasking experiment over a topologically nontrivial Bayesian network. 

To the best of the author's knowledge, \cite{plajnerquestion}, from the field of adaptive testing, was the first and only article to suggest using a quantity like the inferential deviation (in their case ``Expected Skills Variance and Expected Question Variance") for ranking questions (which is analogous to sensor collection ranking). Their variance methods were compared against an information gain; however, the differences between them were found to be statistically insignificant in terms of performance in their experiment. Reference \cite{plajnerquestion} asks for more study in this direction, which we believe we begin to address in the present article by solidifying the foundation of inferential variance, higher order inferential moments, and the inferential probability distribution in Bayesian probability theory.

We recently published conference proceedings titled ``Toward Optimal Conjunction-Based Sensor Tasking using Inferential Moments" \cite{vanslette2023conjunction} based on a preprint of the present article \cite{vanslette2023inferential}, which applied inferential moments for optimal sensor tasking for conjunction events in the domain of space situational awareness. Of interest to the current article, it was proven that optimal sensor tasking using inferential moments is an NP-hard combinatorial optimization problem due to it involving the size-constrained maximization of non-decreasing supermodular objective function. Further, in the Kalman filter paradigm, it was proven that using the mutual information as a sensor tasking objective function is also NP-hard for the same reasons. The sensor tasking experiment in \cite{vanslette2023conjunction} involved choosing to measure one satellite from a potential conjuncting pair (two satellites or resident space objects within some predefined proximity) or another satellite from another potentially conjuncting pair. The sensor tasking experiment was only ``one level deep" in this sense and was computable using brute force and Monte Carlo integration. The main result was that the inferential variance approach lead to 100\% optimal sensor tasking in a mean squared error sense as compared tasking with the mutual information, which only achieved an 88\% success rate.

The remainder of this article is structured as follows: First we will introduce relevant aspects of Bayesian inference in the background section. This is followed by theoretical results, applications, and finally conclusions. The primary results of the present article are theoretical in nature. We apply the theory to a probability distribution described by a Bayesian network to demonstrate concepts in the application section.


\section{Background}
We review notation and relevant aspects of Bayesian inference for reference later in the article.
\subsection{Notation}
Throughout this paper we will be working with a set $U$ of discrete random variables with subsets $A$, $B$, $C$, etc. We denote generic variable elements of these sets using their respective lower case characters $a\in A$ and reserve ``primed" characters $a'\in A'=A$ to represent a particular measured (observed, realized, fixed) element of the set. We will use the three horizontal bars symbol, $\equiv$, in place of an equal sign to mean ``is defined as" when introducing new definitions inline. We will use the starred arrow symbol, $\stackrel{*}{\rightarrow}$, to denote a Bayesian update from a prior distribution to a posterior probability distribution. We use the shorthand $p(a)\equiv p(A=a)$ to denote the probability of $a$ from set $A$ and $p(a|b) \equiv p(A=a|B=b)$ to denote the conditional probability of $a$ given $b$. Further, we will assume all elements are mutually exclusive. We use the standard notation for conditional expectation, $${E}_{X}\Big[f(x)\Big|y\Big] \equiv \sum_{x\in X}f(x)p(x|y).$$
While our results are written in terms of discrete probability distributions, nothing in principle prevents one from making analogous arguments for continuous variables using probability densities or their probability mass functions. Further, the work is not restricted to subsets $A$, $B$, $C$ being single dimensional as straightforward analogous arguments can be made in multidimensional variable settings. 

\subsection{Bayesian Inference -- special cases and applications}
The primary mechanics of Bayesian probability theory are the product and sum probability rules. Bayesian inference uses these mechanics to draw conclusions about the world \cite{Jaynesbook,Catichabook, Cox}. The product rule is the solution for how probability distributions factor under the logical operation of conjunction (and) $p(a,b)=p(a|b)p(b)=p(a)p(b|a)$. Bayes' Rule states that one may update the probability distribution of one variable, $p(a)$, in response to the information that another variable takes a definite value, in this case, $B=b'$, by conditioning on this new information. That is, the Bayesian update from the prior distribution of $a$ to the posterior distribution of $a$ is, 
\begin{eqnarray}
p(a)\stackrel{*}{\rightarrow} p'(a)\equiv p(a|b').\label{bayes}
\end{eqnarray} 
We follow \cite{Catichabook, caticha2006updating, AdomGiffin, vanslette2018quantum} an call this updating rule ``Bayes' Rule" while letting ``Bayes' Theorem" be the relationship among conditional probability distributions,
\begin{eqnarray}
p(a|b')=\frac{p(a,b')}{p(b')}=\frac{p(b'|a)}{p(b')}p(a).\label{bayesTheorem}
\end{eqnarray} 
The sum rule, on the other hand, is the solution for how probability distributions factor under the logical operation of disjunction, that is, the probability $a=1$ or $a=2$ is $p(a=1)+p(a=2)$ (given mutual exclusivity). Marginalization uses the sum rule to quantify the uncertainty of one variable given uncertainty in another variable,
\begin{eqnarray}
p(a)=\sum_{b\in B}p(a|b)p(b).\label{marginalization}
\end{eqnarray} 
These rules are applied across science and industry for reasoning with uncertainties -- sometimes explicitly in terms of probabilities and sometimes implicitly through statistical averages or expectation values. 

We will now review several special cases and formularizations using Bayes' Rule and marginalization for reference later in the article.

\subsubsection{Bayes' Rule as a marginal posterior update}

In the foundational work of \cite{Catichabook, caticha2006updating, AdomGiffin, vanslette2018quantum, vanslette2017entropic}, Bayes' Rule is recovered as a special case of entropic probability updating when the information in question takes the form of data. In this framework, Bayes' Rule takes the form of a marginal posterior, which we will review now. First, let the probability of a variable of interest $a$ be related a variable $b$ through marginalization (\ref{marginalization}),
\begin{eqnarray}
p(a)=\sum_{b\in B}p(a|b)p(b).\nonumber
\end{eqnarray} 
Next, consider $b$ is measured with complete certainty so its value is known to be $b'$. In such a case, the distribution for $b$ can be updated using the standard formulation of Bayes' Rule, 
\begin{eqnarray}
p(b)\stackrel{*}{\rightarrow} p'(b)= p(b|b') = \delta_{b,b'}, \label{data}
\end{eqnarray} 
where $\delta_{b,b'}$ is the Kronecker delta function\footnote{If $b$ is a continuous variable rather than a discrete variable the Kronecker delta is replaced by a Dirac delta function and probabilities are replaced by probability densities.}, which is $1$ if $b=b'$ and is zero otherwise. Using the Kronecker delta in this way represents complete certainty in the value of $b$, which we call a ``data-like" inferential update. The updated knowledge of $b$ leads to an updated knowledge about the value of $a$, which is
\begin{eqnarray}
p(a)=\sum_{b\in B}p(a|b)p(b)\stackrel{*}{\longrightarrow} \sum_{b\in B}p(a|b)p'(b) = \sum_{b\in B}p(a|b)\delta_{b,b'} = p(a|b')= p'(a),\label{marginalPosteriorBayes}
\end{eqnarray} 
which is Bayes' Rule (\ref{bayes}), but from a marginal posterior perspective. 

\subsubsection{Uncertainty Propagation through a Deterministic Function}
Uncertainty propagation is an application of marginalization used to understand or estimate the probability or distributional parameters of one variable given the distribution of another variable and a knowledge of the conditional probabilistic or functional relationship between them. Uncertainty propagation may be thought of as the movement of probability from $p(b)$ to the probability of the variable of interest $p(a)$ through $p(a|b)$.

When relationships between variables are specified by a deterministic functional relationship, e.g. $f:B\rightarrow A$ such that $a=f(b)$, this information can be encoded as a conditional probability that represents certainty \cite{au1999transforming}, i.e., in the discrete case,
\begin{eqnarray}
p(a|b) = \delta_{a,f(b)},\label{determfuncprob}
\end{eqnarray}
again using the Kronecker delta, but this time potentially with real valued indices (if $f(b)$ is a real valued function). Using this conditional probability, the uncertainty from $b$ can be propagated to $a$ theoretically using marginalization
\begin{eqnarray}
p(a) = \sum_{b\in B} p(a|b)p(b) = \sum_{b\in B} \delta_{a,f(b)}\, p(b) = \sum_{b \in B_a} p(b),\label{determfunctionrule}
\end{eqnarray}
where $B_a\subset B$ is the subset of $b$'s that evaluate the value $a$ when passed through $f(b)$. If the function $f$ is uniquely invertible, then $B_a$ only ever contains a single element of $b$, but if non-monotonic, e.g. $a=f(b)=b^2$, then, both $b\pm1$ evaluate to $a=(\pm1)^2=1$ and sensibly, their probabilities sum $p(a=1) = p(b=-1)+p(b=1)$ in equation (\ref{determfunctionrule}).

\section{Theoretical Results: Inferential Theory}

We formulate the theoretical aspects of inferential moments and inferential distributions. We also explore the data-like inferential update special case and a few inferential variance inequalities for applications in the next section. We also derive a power series expansion of the mutual information in terms of inferential moments.
\subsection{Inferential Moments}

Marginalization is widely recognized as the expected value of a conditional probability,
\begin{eqnarray}
p(a)=\sum_{b\in B}p(a|b)p(b)\equiv{E}_B\Big[p(a|b)\Big],\label{expected_marginal}
\end{eqnarray}
which is sometimes noted as such in pedagogical literature, almost as an afterthought. We interpret this equality from a Bayesian inference perspective. If the goal is to infer the value of $a$ from knowledge of $b$, but the value $b$ is uncertain and distributed according to $p(b)$, there is a sense in which the posterior distribution probability value is unknown prior to ``measurement", i.e.,
\begin{eqnarray}
p(a)\stackrel{?}{\rightarrow}p'(a)= p(a|b'=?)=\,?\label{bayesrandom}
\end{eqnarray} 
Thus, before it is observed, $b$ may be considered a random variable and $p(a|b)$, being a function of a random variable $b$, is therefore also a random variable. Not knowing which value of $p(a|b)$ best describes the state of knowledge of $a$ prior to measurement, (\ref{expected_marginal}) can be thought of as an expectation value estimate of the possible \emph{inferences} in (\ref{bayesrandom}). Thus, we can interpret ${E}_B\Big[p(a|b)\Big]=p(a)$ as a sort of \emph{inferential expectation}, which is later recognized as the first order \emph{inferential moment}. Given the fundamental importance of marginalization in Bayesian probability and information theory, we consider formulating an interpretation for higher order conditional probability moments -- which we call inferential moments.

\begin{figure}
\centering
\includegraphics[width = 1\textwidth]{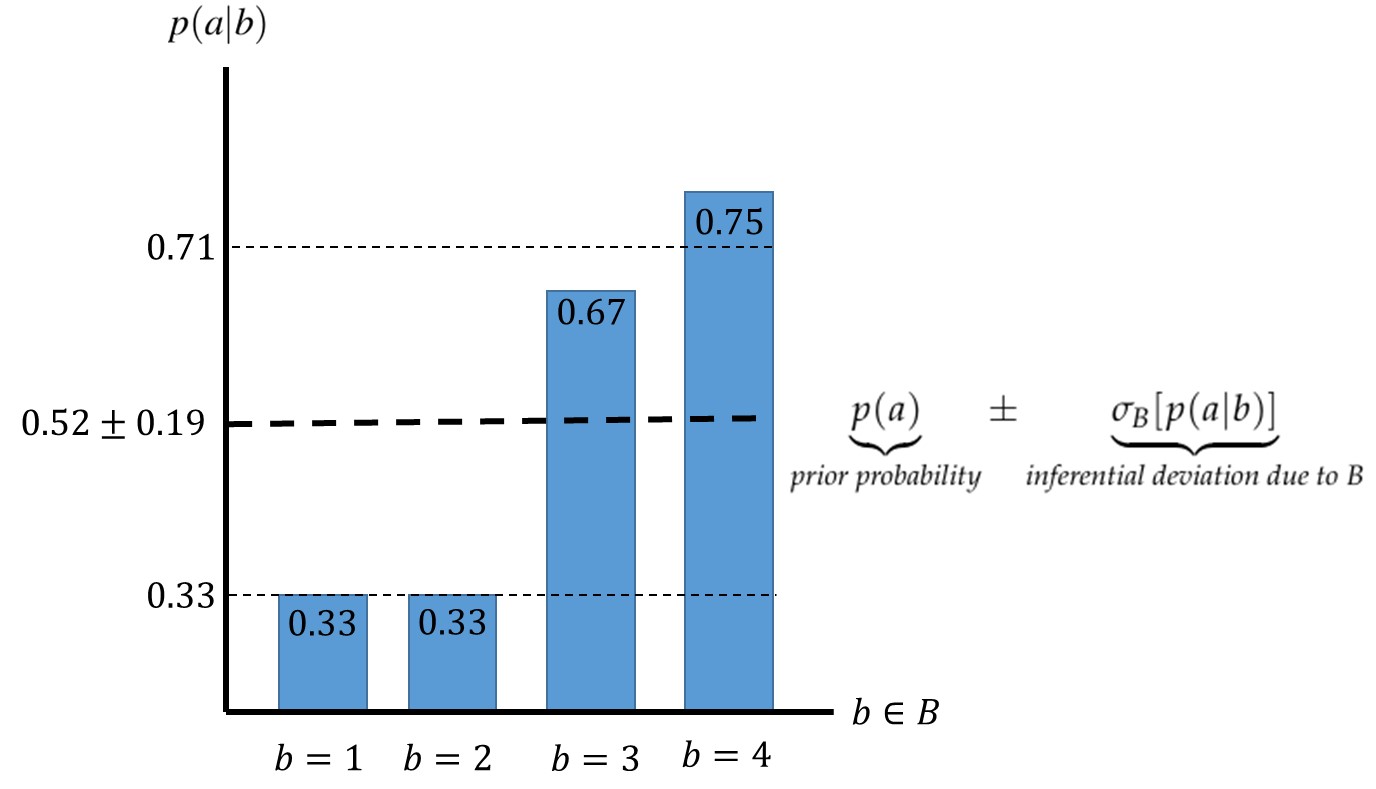}
\caption{ A depiction of an example conditional distribution $p(a|b)$ for a few values of $b$. Further, let the probability any one of the $b$ values to be $p(b)=0.25$ (not depicted). From this information, the prior probability or inferential expectation is computed to be $p(a)=0.52$ and the inferential deviation is $\sigma_B[p(a|b)]=0.19$ which shows the expected deviation of the probability value $p(a)$ from an inferential update of $B$ in (\ref{bayesrandom}). The one sigma band $p(a)\pm \sigma_B[p(a|b)]$ is plotted. }
\label{fig:infvar}
\end{figure}

We find that the second order central inferential moment, the \emph{inferential variance}, can be understood by considering a particular mean square error (${MSE}$) that can be computed in an inference setting. Consider two variables $a$ and $b$ that are known to be related via the joint probability distribution $p(a,b)$. Due to the uncertainty $p(b)$, what is the ${MSE}$ between the canonical estimate for the distribution of $a$, $p(a)$, obtained through marginalization, and the actual distribution of $a$ if $b$ were known, i.e. $p(a|b)$? The ${MSE}$ in this case has target value $\theta = p(a|b)$ and estimated value $\hat{\theta}=p(a) = {E}_B\Big[p(a|b)\Big]$ and is,
\begin{eqnarray}
{MSE}(\theta,\hat{\theta})={E}\Big[(\theta-\hat{\theta})^2\Big]={E}_B\Bigg[\Big(p(a|b) - p(a)\Big)^2\Bigg] ={Var}_B\Big[p(a|b)\Big]\equiv \sigma^2_B\Big[p(a|b)\Big].\label{inferential_variance}
\end{eqnarray}
While the inferential expectation $p(a) = {E}_B\Big[p(a|b)\Big]$ indicates the expected value of $p(a|b)$ when $b$ is uncertain, the second order central moment ${Var}_B\Big[p(a|b)\Big]$ provides a quantification of the expected square error of using $p(a)$ to estimate $p(a|b)$ when $b$ is uncertain, i.e. $p(a)\pm\sigma_B\Big[p(a|b)\Big]$, as depicted in Figures \ref{fig:infvar} and \ref{fig:fig1}. The equalities in (\ref{inferential_variance}) demonstrate the sense in which a quantification of error of a current assessment on the left hand side can also be used to express a predictive uncertainty on the right hand side, i.e. that error and variance are intimately related. Thus, from an inference perspective, ${Var}_B\Big[p(a|b)\Big]$ is the expected variation (or fluctuation) of the probability of one variable in response to an inferential update of another when using Bayes' Rule (\ref{bayesrandom}). For this reason, we will name ${Var}_B\Big[p(a|b)\Big]$ the \emph{inferential variance} and its square root, $\sigma_B\Big[p(a|b)\Big]$, the \emph{inferential deviation}. 

Higher order central inferential moments are computable and give nontrivial results for any suitably behaved joint probability distribution with dependent variables. The interpretation of the $n$th central inferential moments, i.e.,
$${E}_B\Big[(p(a|b)-p(a))^n\Big],$$
such as the \emph{inferential skew}, and multivariate moments, such as the \emph{inferential covariance}, inherit analogous interpretations in this context -- they quantify shape information about the Bayesian probability updating process. If the variables in question are independent, $p(a,b)=p(a)p(b)$, then the inferential variance, as well as all other central inferential moments, are zero, ${E}_B\Big[(p(a|b)-p(a))^n\Big]\rightarrow{E}_B\Big[(p(a)-p(a))^n\Big]=0$. Thus, inferential moments are induced by the marginalization process due to the dependencies between variables of the joint probability distribution. This indicates a potential relationship between inferential moments and mutual information, which is also zero for probabilities that are independent, which we explore a bit further in Section \ref{MIsection}. It should be noted that central inferential moments have a computational complexity that is roughly equal to the cost of performing marginalization (\ref{expected_marginal}) twice -- the first marginalization cost is computing $p(a)$ (taking the expectation of $p(a|b)$) and the second in taking the expectation over the function $(p(a|b)-p(a))^n$ where $p(a)$ is treated as a constant.
%
A key differentiator inferential moments have from the Dirichlet based probabilistic modeling used for subjective logic \cite{subjectivelogic} is that no assumptions other than what is encoded in a given joint probability distribution are needed to perform this analysis. Nothing in principle, however, prevents one from considering inferential moment logic over Dirichlet based distributions. 
\subsection{Inferential Probability Distribution}
In the previous section we introduced the concept of inferential moments, which were derived from treating conditional probabilities as random variables in the inference process. In this inference setting, due to the uncertainty in $B$, conditional probabilities variables are random variables, $p(a|b'=?)$, but, due to Bayes' rule, this implies equally well that the posterior probability variables, $p'(a)$, are random too (\ref{bayesrandom}). A question of interest is ``What is the probability distribution of these random posterior probability variables?" We construct the \emph{inferential probability distribution}, whose moments are equal to the inferential moments in question and which describes the probability distribution of possible posterior Bayesian inference updates due to uncertainty in $B$. Let the ``inferential" probability (mass) distribution be denoted as $\mathcal{P}_B[p'(a)]$ for readability purposes, that’s argument is the posterior probability value $p'(a)\in[0,1]$. We will show that possible posterior probability values $p'(a)$, due to uncertainty in $p(b)$, are distributed uniquely according to the inferential probability distribution, $\mathcal{P}_B[p'(a)].$ 

\begin{figure}
\centering
\includegraphics[width = 0.9\textwidth]{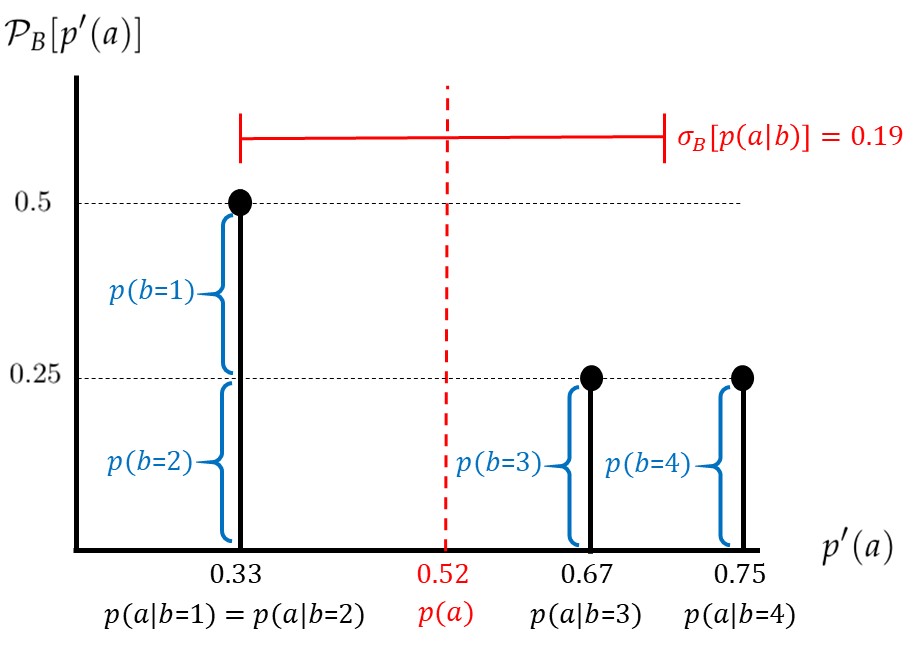}
\caption{An example inferential distribution $\mathcal{P}_B[p'(a)]$ is plotted as the lollipop chart in black against its argument $p'(a)$. The inferential distribution example is calculated from a joint probability distribution with $p(b$=$1)=...=p(b$=$4)=0.25$ and $p(a|b$=$1)=p(a|b$=$2)=0.33$, $p(a|b$=$3)=0.67$, and $p(a|b$=$4)=0.75$. The blue curly braces represent the values of $p(b)$ and how they contribute to the probability value of $\mathcal{P}_B[p'(a)]$ at $p'(a)$ through (\ref{inferential_distribution}). Because $p'(a)$ can equal exactly $0.33$ in two ways ($b=1,2$), $\mathcal{P}_B[p'(a)=0.33] = p(b$=$1) + p(b$=$2) = 0.5$; however, $p'(a)$ can only equal $0.67$ or $0.75$ in one way, so its value is simply the probability of $b$, i.e. $\mathcal{P}_B[p'(a)=0.67] = p(b$=$3)=0.25$ and $\mathcal{P}_B[p'(a)=0.75] =p(b$=$4) = 0.25$. $\mathcal{P}_B$ is zero for all other values of $p'(a)$. The inferential expectation, $p(a)$, is the prior value and the inferential deviation, $\sigma_B\Big[p(a|b)\Big]$, of $p(a,b)$ (or equivalently the mean and standard deviation of $\mathcal{P}_B[p'(a)]$) are plotted in red. This plot shows the sense in which the estimate of the probability of $a$ given unknown $b$ is $0.52\pm 0.19$ similar to Figure \ref{fig:infvar}; however, the exact distribution of posterior probabilities is quantifiable and given by $\mathcal{P}_B[p'(a)]$, as shown. }
\label{fig:fig1}
\end{figure}

To construct $\mathcal{P}_B[p'(a)]$, we will use the ideas from equations (\ref{determfuncprob}) and (\ref{determfunctionrule}). In our problem, we know the deterministic function that relates conditional probability values $p(a|b)$ to posterior probability values $p'(a)$ -- it is simply Bayes' rule $p'(a)=f_a(b) = p(a|b)$. Thus, given a value $b$, the posterior distribution $p'(a)$ is known with certainty, so we can use (\ref{determfuncprob}) and equate the conditional inferential distribution with the Kronecker delta $\mathcal{P}_B[p'(a)|b] = \delta_{p'(a),p(a|b)}$, which expresses the deterministic map $p'(a)=f_a(b) = p(a|b)$ with certainty. We then use (\ref{determfunctionrule}) to constructed the inferential probability distribution,
\begin{eqnarray}
\mathcal{P}_B[p'(a)] = \sum_{b\in B} \mathcal{P}_B[p'(a)|b]\, p(b) = \sum_{b\in B} \delta_{p'(a),p(a|b)}\, p(b) = \sum_{b \in B_{p'(a)}} p(b). \label{inferential_distribution}
\end{eqnarray}
An example of how $\mathcal{P}_B[p'(a)]$ is constructed from (\ref{inferential_distribution}) is depicted in Figure \ref{fig:fig1} along with its first and second (inferential) moments. This is the same example probability distribution $p(a,b)$ from Figure \ref{fig:infvar}.


The moments of the inferential distribution are equal to the inferential moments of the joint distribution in question, as can be seen from the moment generating function $\mathcal{M}\Big[...\Big]$ of $\mathcal{P}_B[p'(a)]$,
\begin{eqnarray}
\mathcal{M}\Big[\mathcal{P}_B[p'(a)]\Big] &=& \sum_{p'(a)}\exp\Big(t \,p'(a)\Big)\mathcal{P}_B[p'(a)]\nonumber\\
&=& \sum_{p'(a)}\exp\Big(t \,p'(a)\Big)\cdot \sum_{b\in B} \delta_{p'(a),p(a|b)} p(b)\nonumber\\
&=& \sum_{b\in B}\exp\Big(t \,p(a|b)\Big) p(b).\label{MGF}
\end{eqnarray}
Explicitly, the moment generating function relationship indicates the inferential distribution has moments equal to the inferential moments of the distribution in question, which can be seen using the sifting property of the Kronecker delta,
\begin{eqnarray}
{E}_{\mathcal{P}_B}\Big[p'(a)^n\Big] \equiv \sum_{p'(a)}p'(a)^n\,\mathcal{P}_B[p'(a)] = \sum_{p'(a)}p'(a)^n\sum_{b\in B} \delta_{p'(a),p(a|b)} p(b) = {E}_{B}\Big[p(a|b)^n\Big],
\end{eqnarray}
and therefore it also has central moments that are equal to the central inferential moments in question,
\begin{eqnarray}
{E}_{\mathcal{P}_B}\Bigg[\Big(p'(a)-{E}_{\mathcal{P}_B}\Big[p'(a)\Big]\Big)^n\Bigg]={E}_B\Big[(p(a|b)-p(a))^n\Big].
\end{eqnarray}
Due to the moment generating function relationship in (\ref{MGF}), which holds for all $t$, the inferential probability distribution is the unique distribution for representing uncertainty in the posterior values of a distribution that could or will experience a Bayesian update,
\begin{eqnarray}
p(a)\stackrel{*}{\rightarrow} p'(a)\sim \mathcal{P}_B[p'(a)].
\end{eqnarray}
That is, when information about dependent variables is unknown, unavailable, or yet to be observed, Bayesian probability updating can be represented as a random process having random variable $ p'(a)\sim \mathcal{P}_B[p'(a)]$. This analysis uncovers intrinsic probabilistic structure inherent in the process of probability updating and joint probabilities themselves. 

By nature of being a probability distribution, the inferential probability distribution allows one to answer more pinpointed questions not immediately answerable with moment information alone. Using the inferential probability distribution above as an example, one can directly answer interval-like questions, for instance: what is the probability the posterior distribution $p'(a)$ will take a probability value greater than $0.52$, i.e. $ \mathcal{P}_B[p'(a)\,$>$\,0.52] = 0.5$, or what is the probability the posterior distribution is $0.33$, which is $ \mathcal{P}_B[p'(a)\,$=$\,0.33] = 0.5$, or the probability the posterior distribution will be within one standard deviation, i.e. $\mathcal{P}_B[0.33\,$$\leq$$ \,p'(a)\,$$\leq$$\,0.71] = 0.75$. This sort of investigation could be useful in alert-like applications that make alerts if a probability value is greater than some threshold for hidden Markov type models.

\subsection{Data-like inferential updates and the Bernoulli distribution as a special case}

Although the inferential moment and inferential probability distribution analysis presented thus far have been framed in terms of joint probability distributions, we make a point to state that this logic also apply to probability distributions over single variable distributions $p(a)$ when there is an implicit assumption that the variables $a$ can be measured directly and updated via Bayes, i.e. the data-like inferential update in (\ref{data}). Letting $A'=A$ be measurement variables such that $p(a|a')=\delta_{a,a'}$ and $p(a')=p(a)$ for $a'=a$, the $n$th central moment in this case is $${E}_{A'}\Big[(\delta_{a,a'}-p(a))^n\Big]= \sum_{a'\in A'}p(a')\Big(\delta_{a,a'}-p(a)\Big)^n.$$ The inferential variance in this case is simply ${E}_{A'}\Big[(\delta_{a,a'}-p(a))^2\Big]=p(a)-p(a)^2$, which we note is equal to the variance of a Bernoulli variable given by $p(a)=p$. This relationship ends up holding for all $n$ as the moment generating function of the inferential probability distribution in this case is,
\begin{eqnarray}
\mathcal{M}\Big[\mathcal{P}_{A'}[p'(a)]\Big] &=&\sum_{a'\in A'}\exp\Big(t \,p(a|a')\Big) p(a')\nonumber\\
&=&\sum_{a'\in A'}\exp\Big(t \,\delta_{a,a'}\Big) p(a')\nonumber\\
&=&\exp\Big(t\cdot 1\Big) p(a)+\sum_{a' \neq a}\exp\Big(t\cdot 0\Big) p(a')\nonumber\\
&=&\exp(t) p(a)+(1-p(a))\nonumber\\
&=&\exp(t) p+q,
\end{eqnarray}
which is the moment generating function of the Bernoulli distribution with $p=p(a)$ and $q=1-p(a)$. Shown directly, under a data-like update the inferential probability distribution is, 
\begin{eqnarray}
\mathcal{P}_{A'}[p'(a)] &=& \sum_{a'\in A'} \delta_{p'(a),p(a|a')}\, p(a') \nonumber\\
&=& \sum_{a'\in A'} \delta_{p'(a),\delta_{a,a'}}\cdot p(a') \nonumber\\
&=& \delta_{p'(a),1}\cdot p(a) + \sum_{a'\neq a} \delta_{p'(a),0}\cdot p(a') \nonumber\\
&=& \delta_{p'(a),1}\cdot p(a) + \delta_{p'(a),0}\cdot (1-p(a)) \nonumber\\
&=& p(a)^{\delta_{p'(a),1}}\cdot(1-p(a))^{1-\delta_{p'(a),0}} \nonumber\\
&=& p^{x}q^{1-x},
\end{eqnarray}
which is again the Bernoulli distribution with $p=p(a)$, $q=1-p(a)$, and $\mathcal{P}_{A'}[p'(a)]$ only taking nonzero probabilities for $x=p'(a)\in\{0,1\}$. Thus, in the special case when the inferential update is data-like, the inferential probability distribution is Bernoulli distributed. Graphically, the inferential distribution in this case has lollipop sticks on the edges as $p'(a)$ is equal to 0 and 1 and with inferential probability values (i.e. lollipop height) given by $p$ and $q$ respectively.

\subsection{Inferentially Updating Inferential Moments and the Inferential Probability Distribution}

We can update inferential moments and the inferential probability distribution in response to new information using the marginal posterior Bayes' rule (\ref{marginalPosteriorBayes}) with (\ref{expected_marginal}). We summarize the results of the proofs from Appendix \ref{appendix1} in this subsection.

Due to (\ref{expected_marginal}), the marginal posterior Bayes' rule (\ref{marginalPosteriorBayes}) can instead be written as a Bayes' rule update of the inferential expectation. That is, $p(a)\stackrel{*}{\rightarrow}p'(a)$ is equally written as,
\begin{eqnarray}
p(a) = {E}_{B}\Big[p(a|b)\Big]\stackrel{*}{\rightarrow}{E}_{B}\Big[p(a|b)\Big|b'\Big]=p'(a).\label{bayes2text}
\end{eqnarray}
We use these relationships to argue for how higher order central inferential moments update. We find that the posterior central inferential moments are zero when the uncertain variable in question has been ``completely" measured,
\begin{eqnarray}
{E}_{B}\Big[(p(a|b)-p(a))^n\Big] \stackrel{*}{\rightarrow} {E}_{B}\Big[\Big(p(a|b)-p'(a)\Big)^n\Big|b'\Big]= 0,
\end{eqnarray}
because the posterior inferential probability distribution is singular,
\begin{eqnarray}
\mathcal{P}_B[p'(a)]\stackrel{*}{\rightarrow}\mathcal{P}_B[p'(a)|b'] = \delta_{p'(a),p(a|b')}.
\end{eqnarray}
If instead there are more than one measurable variables under consideration in the inference problem (i.e. $B$ and $C$) and one is measured, this results in the following posterior inferential moments:
\begin{eqnarray}
{E}_{B,C}\Big[(p(a|b,c)-p(a))^n\Big] \stackrel{B}{\rightarrow} {E}_{C}\Big[\Big(p(a|b',c)-p(a|b')\Big)^n\Big|b'\Big],\label{reduceupdateruleintext}
\end{eqnarray}
which correspond to the posterior inferential probability distribution,
\begin{eqnarray}
\mathcal{P}_{B,C}[p'(a)]\stackrel{B}{\rightarrow}\mathcal{P}_{B,C}[p'(a)|b'] =\sum_{c\in C} \delta_{p'(a),p(a|b',c)}\, p(c|b').\label{inferential_prob_update}
\end{eqnarray}
The prior inferential variance is updated to a posterior inferential variance,
\begin{eqnarray}
{Var}_{B,C}\Big[p(a|b,c)\Big]\stackrel{B}{\rightarrow}{Var}_{C}\Big[p(a|b',c)\Big|b'\Big].\label{var_bayes}
\end{eqnarray}

\subsection{Inferential Variance Inequalities}
Appendix \ref{appendix1} further derives the following inferential variance equality, which we call the inferential variance decomposition formula:
\begin{eqnarray}
{Var}_{B,C}\Big[p(a|b,c)\Big]={E}_B\Bigg[{Var}_{C}\Big[p(a|b,c)\Big|b\Big]\Bigg]+{Var}_{B}\Big[p(a|b)\Big],\label{total_inferential_variance}
\end{eqnarray}
which decomposes the ``total" or ``multivariable" inferential variance into into the expected posterior inferential variance and the \emph{partial} inferential variance, respectively. Because all of the terms in the above equation are greater than or equal to zero, it implies the following inequalities: First that the partial inferential variance is less than the total,
\begin{eqnarray}
{Var}_{B}\Big[p(a|b)\Big]\leq {Var}_{B,C}\Big[p(a|b,c)\Big],\label{partial_inequality}
\end{eqnarray}
and secondly that the \emph{expected} posterior inferential variance is less than or equal to the total inferential variance,
\begin{eqnarray}
{E}_B\Bigg[{Var}_{C}\Big[p(a|b,c)\Big|b\Big]\Bigg]\leq{Var}_{B,C}\Big[p(a|b,c)\Big]\label{posterior_inequality}.
\end{eqnarray}
Equation (\ref{total_inferential_variance}) is an odd form of the law of total variance under special conditions for these probability variables, which is derived at the end of the Appendix \ref{appendix1}.

\subsection{Exhaustive Inferential Variance}
While we will not need the exhaustive inferential variance in this article, we construct it for completeness. So far, we have only considered the inferential moments and inferential variance of a single argument $a\in A$ due to uncertainties in other variables. We suggest the following measure to quantify the inferential variance exhaustively over the entire distribution, which we call the exhaustive inferential variance, 
\begin{eqnarray}
\bigtriangleup^2[A]_B = \sum_{a\in A} {Var}_{B}\Big[p(a|b)\Big]\geq 0.
\end{eqnarray}
We find this measure preferable for this purpose as compared to one that accounts for inter-element correlations, i.e. summing over $a$ inside rather than outside the variance, which is ${Var}_{B}\Big[\sum_a p(a|b)\Big] = {Var}_{B}[1] = 0$ due to normalization, or said otherwise, due to probability conservation.

If the sum over $a$ was instead an integral, the exhaustive inferential variance almost starts to resemble a 2-Wasserstein distance $W_2$, but the quantities are distinct. The 2-Wasserstein distance has explicit dependence on the spatial distance between distributions as can be seen in the Wasserstein distance between point distributions $W_2\Big[\delta_{x,x_0},\delta_{x,x_1}\Big] = ||x_0-x_1||_2$, i.e. it is the distance between coordinate values. The exhaustive inferential variance instead is all about measuring how much the probability distribution of each argument is expected to vary in response to an inference, independent of spatial considerations. 

In \cite{vanslette2023conjunction} the exhaustive inferential variance in the continuous domain, $\bigtriangleup^2(X)_Y \equiv \int_{x\in X}{Var}_Y\Big[\rho(x|y)\Big]\,dx\geq 0$, was compared to the MI for multivariate joint Gaussian variables under a Kalman-filter measurement update. We state those results as an example. Given the conditional multivariate Gaussian is,
\begin{eqnarray}
\rho(x|y) = N\Big(\mu_X'(y),\Sigma_{X}'\Big)= N\Big(\mu_X+\Sigma_{XY}\Sigma_{Y}^{-1}(y-\mu_Y),\Sigma_X-\Sigma_{XY}^T\Sigma_{Y} ^{-1}\Sigma_{XY}\Big),
\end{eqnarray}
the exhaustive inferential variance is,
\begin{eqnarray}
\bigtriangleup^2(X)_Y &=&\int_{x,y\in X,Y}N\Big(\mu_X'(y),\Sigma_{X}'\Big)^2\rho(y)\,dy\,dx-\int_{x\in X}N\Big(\mu_X,\Sigma_{X}\Big)^2\,dx\nonumber\\
&=& \frac{1}{\sqrt{2}(2\pi)^{k_X/2}}\Big(\frac{1}{|\Sigma_X'|^{\frac{1}{2}}}-\frac{1}{|\Sigma_X|^{\frac{1}{2}}}\Big),
\end{eqnarray}
which shares some features of the MI in this case \cite{Adurthi,schmaedeke1993information},
\begin{eqnarray}
\mbox{MI}[X;Y]&=&\frac{1}{2}\ln\Big(\frac{|\Sigma_X|}{|\Sigma_X'|}\Big).
\end{eqnarray}

\subsection{Inferential Moments and Mutual Information\label{MIsection}}
%

We find a power series expansion of the MI in terms of inferential moments. First rewrite the MI like,
\begin{eqnarray}
{MI}[A,B] &=& \sum_{b}p(b)\sum_a\Bigg[p(a|b)\ln\Big(p(a|b)\Big)\Bigg] -\sum_ap(a)\ln\Big(p(a)\Big),\label{MI2}
\end{eqnarray}
then consider substituting a power series expansion of the expression $p(a|b)\ln\Big(p(a|b)\Big)$. This expression is a function of $p(a|b)$, which we will expand about the constant $p(a)$. The power series, $f(x)=\sum_{n=0}^{\infty}\frac{f^{(n)}(x_0)}{n!}(x-x_0)^n$, with $x=p(a|b)$, $x_0=p(a)$, and $f(x)=p(a|b)\ln(p(a|b))$, yields,
\begin{eqnarray}
p(a|b)\ln\Big(p(a|b)\Big) &=& p(a)\ln\Big(p(a)\Big) + \Big(1+\ln(p(a))\Big)(p(a|b) - p(a)) \nonumber\\
&+& \frac{1}{2p(a)}(p(a|b)-p(a))^2 - \frac{1}{6 p(a)^2}(p(a|b)-p(a))^3+ ... \nonumber\\
&=& p(a)\ln\Big(p(a)\Big) + \Big(1+\ln(p(a))\Big)(p(a|b) - p(a)) \nonumber\\
&+& \sum_{n=2}^{\infty} \Big(\frac{(-1)^n}{ n(n-1)\cdot p(a)^{n-1}}\Big)(p(a|b)-p(a))^n.
\end{eqnarray}
The radius of convergence of this series is $r_a=\lim_{n\rightarrow\infty}\Big|\frac{c_n}{c_{n+1}}\Big| = p(a)$, so the series converges absolutely and uniformly on compact sets inside the open disc defined by $r_a$. That is, the series converges when all $p(a|b)$ are ``close enough" in value to their respective $p(a)$, i.e. $|p(a|b)-p(a)|\leq r_a=p(a)$. Stated informally, the series converges when $A$ and $B$ have a weak dependence on one another, which consequently are instances of (relatively) low MI. Substituting this expansion into the (\ref{MI2}), summing over $b$, and subtracting off the 0th order term $p(a)\ln\Big(p(a)\Big)$, yields an expansion of the ${MI}$ in terms of the central inferential moments,
\begin{eqnarray}
{MI}[A,B] &=& \sum_{a} \sum_{n=2}^{\infty} \Big(\frac{(-1)^n}{ n(n-1)\cdot p(a)^{n-1}}\Big)\cdot{E}_B\Big[(p(a|b)-p(a))^n\Big],\label{MI_IM}
\end{eqnarray}
where the first nonzero term is proportional to the inferential variance.\footnote{ The linear term is zero once summing over $b$ because the first order central inferential moment ${E}_B[p(a|b)-p(a)]=0$.} Because odd inferential moments need not be positive, the series is not alternating in general.

While this series representation does not converge for all $p(a,b)$, we believe the connection between the MI and inferential moments is worth making. Along with the relationship to MSE (\ref{inferential_variance}), this connection reinforces the idea from inferential moments might be useful tools for performing inference tasks traditionally performed with information theoretic tools like the MI. We demonstrate this in our applications in Section \ref{applications}.

\section{Applications of Inferential Theoretic Logic\label{applications}}

We utilize the interpretation of inferential deviations as a measure of posterior estimation root mean squared error from (\ref{inferential_variance}) to perform tasks typically handled within information theory. Python 3.8 was used inside a Jupyter Notebook. Network structures were adopted from PyBBN \cite{vang_2017} but new functions were written to support our experiments/applications.


\subsection{Improved Decision Making Under Uncertainty}
Probabilistic graphical models such as Bayesian networks and hidden Markov Models can be thought of as tools that provide probabilistic state estimates for the purpose of situational awareness and downstream decision making. These tools are most useful when applied in situations where the direct observation of a state of interest is not possible and one must rely on making inferences from the observation of ancillary state variables to understand the probability of a state of interest. In computational intelligence systems, probabilistic estimates computed in this way often provide information for downstream decision making. State probabilities of interest are computed by marginalizing over the distributions of ancillary variables (prior or posterior if there are observations).

From the perspective of this article, current probabilistic graphical models only quantify distributional uncertainty using the first order inferential moment, the inferential expectation. We show how decision making can be improved by quantifying the inferential deviations of a Bayesian network or its inferential probability distributions. While information theoretic quantities can make statements about how correlated one variable is to other variables in a Bayesian network, this does not address how much the probability might change in response to new measurements, which is arguably more useful for inference and analysis. 

\begin{figure}
\centering
\includegraphics[width = 1\textwidth]{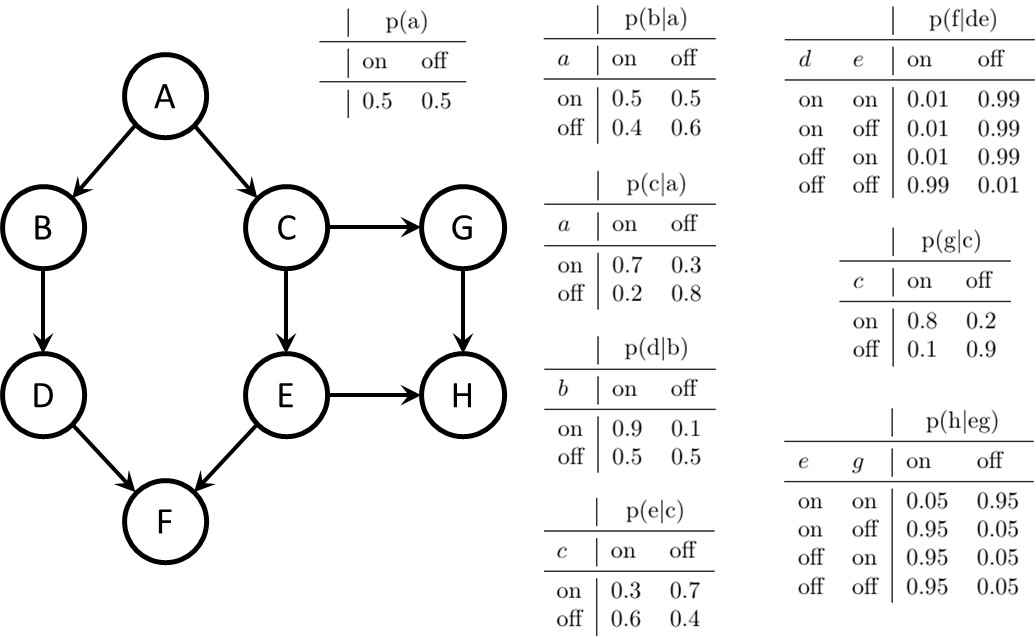}
\caption{The pedagogical Bayesian network example given by \cite{Huang}. Nodes take one of two values -- on or off.}
\label{fig:fig2}
\end{figure}

We apply inferential deviation calculations corresponding to a pedagogical Bayesian network example given by \cite{Huang}, depicted in Figure \ref{fig:fig2}, and record inferential deviation results in Table \ref{table1}. This example was chosen due to it being topologically nontrivial yet simple enough that its binary variables and probability tables can be written out explicitly.

The results of Table \ref{table1} demonstrate the utility of quantifying inferential deviations and we discuss how they can better inform decision making. 
Consider the state of interest is the ``on/off" state of node F. A vanilla Bayesian network will indicate the prior probability of the state of F being ``on" is 18\%, as shown in the first column of the table. Given this probabilistic estimate, a practitioner may find this probability small enough to make a decision or action, perhaps on the basis of a $<$25\% threshold. However, being a prior distribution, no state information has been measured about ancillary variables and the distribution of F has a propensity to change, or said otherwise, there is a sense in which the 18\% is not particularly trustworthy (although it is theoretically the best estimate in a mean squared error sense). By further quantifying the inferential deviation of node F, we see that the node has a relatively large inferential deviation, meaning that if more information was gained, the probability of node F could change considerably. Taking both the prior distribution and its inferential deviation into account, a decision maker would be more hesitant to act on the 18\%$\pm$37\% probability as it is clear there is a good chance the probability could be well above the 25\% threshold. After a measurement of E being ``on" (third column), the probability of F being ``on" is 1\%$\pm$0\%, which indicates there is no more knowledge to gain from ancillary states and which is clearly below the 25\% threshold. If instead E is ``off", the probability F is on is 32\%$\pm$46\%, which while above the 25\% threshold, has considerable inferential variance. The knowledge and quantification of the inferential deviation provided additional information about how the distribution is expected to change, which can lead to improved decision making. 
%
\begin{table}

\begin{center}
\begin{tabular}{c|cccc}
Node X & $p\pm\sigma_{{tot}}[p]$ & $p\pm\sigma_{E}[p]$ & $p'\pm\sigma_{{tot}}\Big[p'\Big|$e\footnotesize =on\normalsize$\Big]$ & $p'\pm\sigma_{{tot}}\Big[p'\Big|$e\footnotesize =off\normalsize$\Big]$ \\
\hline
\,\,\,\,\,\,A & $0.5\pm0.25$ & $0.5\pm0.08$ & $0.42\pm0.23$ & $0.57\pm0.25$ \\
\,\,\,\,\,\,B & $0.45\pm0.22$ & $0.45\pm0.01$ & $0.44\pm0.22$ & $0.46\pm0.22$ \\
\,\,\,\,\,\,C & $0.45\pm0.38$ & $0.45\pm0.15$ & $0.29\pm0.34$ & $0.59\pm0.37$ \\
\,\,\,\,\,\,D & $0.68\pm0.36$ & $0.68\pm0.003$ & $0.68\pm0.20$ & $0.68\pm0.45$ \\
\,\,\,\,\,\,E & $0.46\pm0.37$ && \\
\,\,\,\,\,\,F & $0.18\pm0.37$ & $0.18\pm0.16$ & $0.01\pm0.0$ & $0.32\pm0.46$ \\
\,\,\,\,\,\,G & $0.41\pm0.39$ & $0.41\pm0.10$ & $0.3\pm0.42$ & $0.51\pm0.34$ \\
\,\,\,\,\,\,H & $0.82\pm0.31$ & $0.82\pm0.14$ & $0.68\pm0.41$ & $0.95\pm0.0$ \\
\hline
\end{tabular}
\caption{\label{table1} A partial quantification of the inferential deviations corresponding to the Bayesian network example given by Huang. A node's marginal probability of node being in the ``on" state along with its inferential deviation, $p\pm \sigma[p]$, is tabulated per node ``X" for a few related informational states. Here, the $p\pm\sigma_{{tot}}[p]$ column contains prior total inferential deviations, i.e. before measurements or updates, i.e. $p(x)\pm\sigma_{{tot}}[p(x|$all other nodes$)]$ where $x={on}$. The $p\pm\sigma_{E}[p]$ column is the partial inferential deviation between $p(x|$e$)$ and $p(x)$ for node $X$, i.e. $\sigma_{E}[p(x|e)]$. The $p'\pm\sigma_{{tot}}\Big[p'\Big|$e\footnotesize =on\normalsize$\Big]$ and $p'\pm\sigma_{{tot}}\Big[p'\Big|$e\footnotesize =off\normalsize$\Big]$ \small columns are the $p'=p'(x)=p(x|e)$ posterior marginal distributions (\ref{bayes2text}) with their corresponding posterior total inferential deviations (\ref{var_bayes}) after an update of $E$ being in the ``on" or ``off" state, respectively. Posterior node $E$ values are omitted as updating $E$ with respect to $E$ is no-longer an ancillary variable update. }
\end{center}
\end{table}
Aspects of the inferential variance decomposition formula (\ref{total_inferential_variance}) are apparent Table \ref{table1}. Indeed the partial inferential deviations are less than the total inferential deviations (the second column standard deviations are less than the first column standard deviations). Further, note that the partial inferential deviation, $\sigma_B\Big[p(a|b)\Big]$, is exemplified each row, for example, in row $A$, the posterior probabilities $0.42$ and $0.57$ from measuring E (columns 3 and 4) are about $0.08$ (the partial inferential deviation, column 2) away from the prior $0.5$. While some posterior total inferential deviations increase (the third and fourth column compared to the first), the expected value of the posterior inferential variances satisfy (\ref{posterior_inequality}). 

\begin{figure}
\centering
\includegraphics[width = 0.5\textwidth]{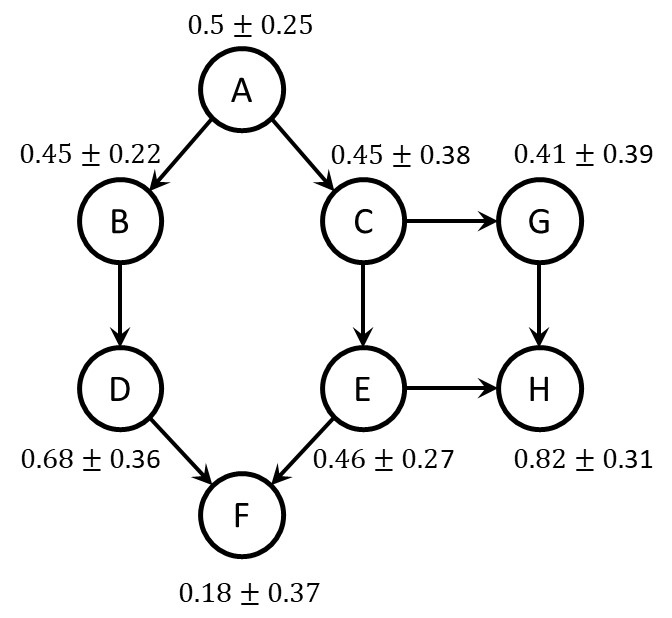}
\caption{The prior state of the Bayesian network with total inferential deviations.}
\label{fig:IVBN}
\end{figure}
The mix of prior probabilities and inferential deviations can provide soft information toward understanding the (in this case the prior) state of a Bayesian network as seen in Figure (\ref{fig:IVBN}). One case easily see that the probability of $B$ is expected to change the least while $G$ is expected to change the most.

\begin{figure}
\centering
\includegraphics[width = .9\textwidth]{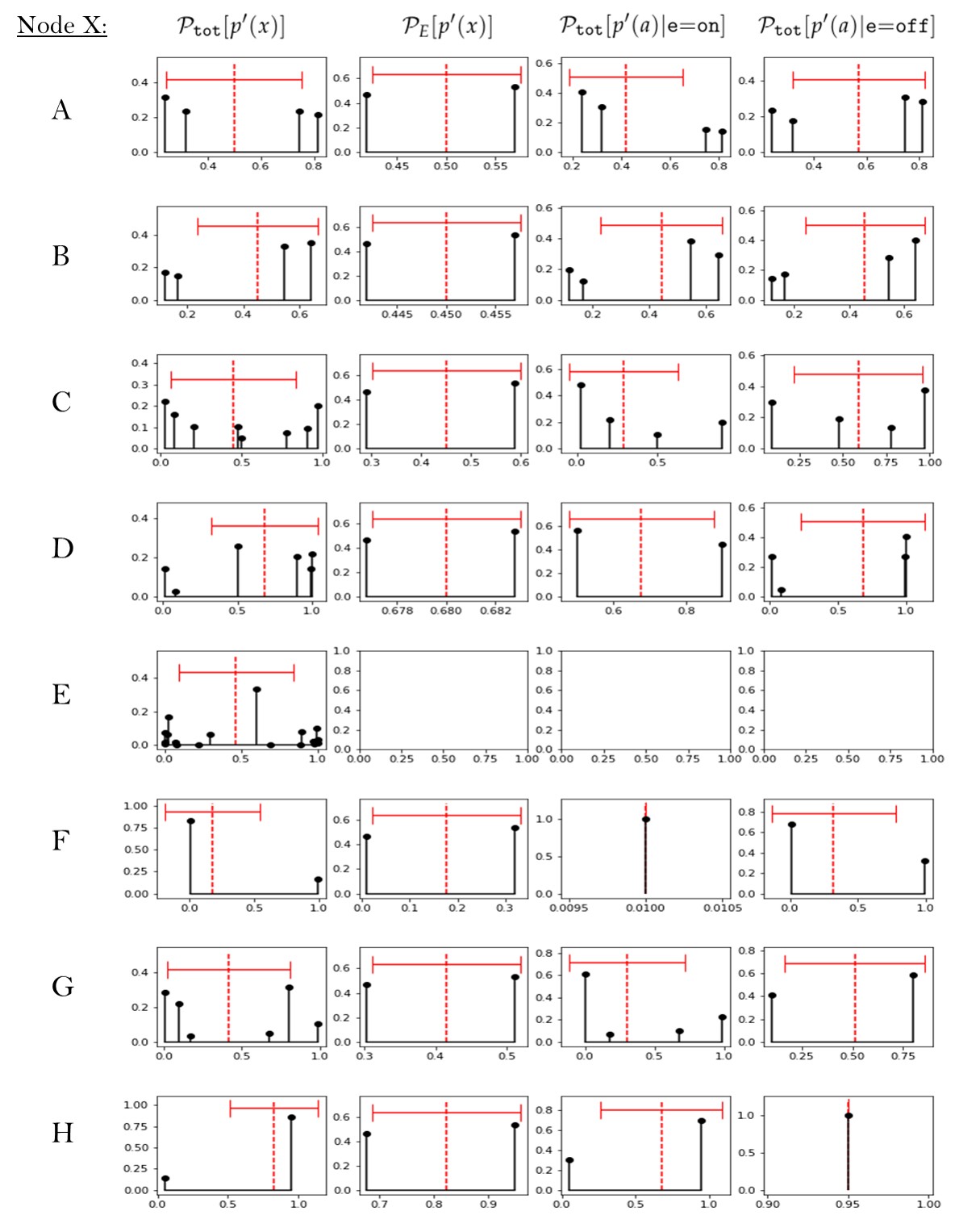}
\caption{The inferential probability distribution corresponding to each entry of Table \ref{table1} is plotted in black along with corresponding means and standard deviations in red, which are equal to the inferential expectation and inferential deviation values reported in Table \ref{table1}. Per plot, the vertical axis is the corresponding inferential probability distribution in the columns heading and the horizontal axis is its corresponding posterior probability argument. The first column heading is $\mathcal{P}_{{tot}}[p'(x)]$ which is the "total" inferential probability distribution corresponding to the node $X$, e.g. for node $A$, it is $\mathcal{P}_{{tot}}[p'(a)]=\mathcal{P}_{BCDEFGH}[p'(a)]$. This is the distribution of posterior values if all nodes except $X$ were measured. The second column heading is $\mathcal{P}_{E}[p'(x)]$ and is the inferential probability distribution of posteriors from measurements of $E$ alone, which is why there is only up to 2 posterior values -- one posterior for $e$=on and one posterior for $e$=off -- and why their inferential probability distributions are similar shapes (\ref{inferential_distribution}). The last two columns are the $E$ conditional total inferential probability distributions (\ref{inferential_prob_update}), e.g. for row $A$, is $\mathcal{P}_{{tot}}[p'(a)|{e}{=}{on}]=\mathcal{P}_{BCD FGH}[p'(a)|{e}{=}{on}]$ and $\mathcal{P}_{{tot}}[p'(a)|{e}{=}{off}]=\mathcal{P}_{BCD FGH}[p'(a)|{e}{=}{off}]$. } 
\label{fig:tablefig}
\end{figure}

 There is an argument that quantifying probabilistic uncertainties using inferential deviations is conceptually simpler than quantifying them with entropy based quantities, which may not be on the same 0 to 1 scale as the probability distribution they are referencing and which have interpretations that differ from probability. Users of probabilistic graphical models can get a better understanding of the state of their system and improve situational awareness by quantifying inferential deviations along with marginal probabilistic inferences. 


Figure \ref{fig:tablefig} corresponds to Table \ref{table1} but is instead calculated using inferential probability distributions. Quantifying possible posteriors in this way allows you to go beyond the analysis presented in Table \ref{table1}. Looking again at node $F$, while the inferential deviation information 18\%$\pm$37\% provides some caution about taking action on a hypothetical 25\% threshold, the inferential probability distribution provides a more precise quantification, i.e. that, prior to measurement, there is an 83\% chance $F$ will yield a posterior probability below the threshold after measuring all ancillary nodes. If $E$ is measured to be off, the updated inferential probability distribution indicates there is now a 68\% chance that the posterior probability of $F$ will be below the 25\% threshold.


\subsection{Inferential Deviations and Toward Optimal Sensor Tasking For Bayesian Networks\label{sensor_tasking}}
We break this material into several subsections: Motiviation, Problem Statement, Algorithms, and Results.
\subsubsection{Motivation}
A problem of interest related to situational awareness is the continual tracking of states that cannot be measured directly through the collection of ancillary measurements. In many settings, it is not feasible to monitor all ancillary variables simultaneously. Thus, the goal is to develop an optimal sensor tasking schedule for the purpose of ``refining" the probability estimate of unknown states of interest. Because inferential moments give information about the expected response of a probability distribution to new information, we can utilize them to rank the expected inferential benefits of utilizing one sensor over another for the purpose of optimal sensor tasking and probabilistic estimate refinement. Decision theoretic utility functions that define the availability of a sensor is outside the scope of this article. 

The tools available in the literature for performing optimal sensor tasking typically seek sensors that maximize the MI, its related quantity, the Information Gain (IG), or the Fisher information (which is a local approximation of the MI). We can ask the question however -- ``For optimal sensor tasking, why would it be preferable to task a sensor according to a maximum MI, i.e. by an amount of information/interdependence between variables, if we could instead task the sensor that is expected to eliminate the maximum amount of epistemic probabilistic estimation error?" It seems extremely reasonable that the preferred goal of sensor tasking in many cases would be to reduce probabilistic state estimation error rather than gaining information to improve state inference, especially given that the later does not imply an optimal reduction of probabilistic estimation error, as we will show.

We propose that using an inferential variance-based sensor tasking objective will rank sensor collections by their expected ability to improve the accuracy of your probabilistic state estimate due to the inferential variance and expected MSE equality in (\ref{inferential_variance}). We propose tasking sensors according to the maximum partial inferential variance using (\ref{partial_inequality}), which is a greedy approach to sensor tasking aimed at reducing expected probabilistic error. We demonstrate that one can use inferential deviations to task sensors to directly reduce the remaining root mean squared probability estimation error, improving over the mutual information based methods.

\subsubsection{Problem Statement}
We consider the following optimal sensor scheduling problem over the Huang network. Let the joint state variables set in question $U$ (e.g $A\times B\times C$) and the complementary joint ancillary variables be $V$ (e.g. $D\times E\times F\times G\times H$). Prior to making measurements the probability estimate of $u$ is the prior distribution $p(u)$. This probability is updated by making measurements and obtaining measured values from the nodes in $V$. This generates "inference trajectories", for example, for $u=c\,$ is ``on", and inference trajectory might be,
\begin{eqnarray}
p(c={on})&\stackrel{G}{\rightarrow}&p(c={on}| g={off})\nonumber\\
p(c={on}| g={off})&\stackrel{A}{\rightarrow}&p(c={on}| a={on}, g={off})\nonumber\\
p(c={on}| a={on}, g={off})&\stackrel{E}{\rightarrow}&p(c={on}| a={on}, e={on}, g={off}),\nonumber
\end{eqnarray}
that is, measuring $G\rightarrow A\rightarrow E$.

We propose that the goal of the optimal sensor scheduling problem is to design an algorithm that generates inference trajectories that most quickly converge probabilistic state estimates to the target underlying probability of interest. Let $p(u|v^{'(n)})$ be the current estimate, where $v^{'(n)}$ be the $n$ of $N$ measured node values of $V^{(n)}\subset V$ and let $p(u|v')$ be the target probability of interest, where $v'\in V$ are the (hidden) definite values of $V$ that will inevitably be revealed sequentially by measurement. We select the ``measure of goodness" of an inference trajectory to be how quickly it converges in a square error sense by considering the sum of the square error between the current estimate and the target, $(p(u|v')-p(u|v^{'(n)}))^2$, at every step of the inference trajectory, i.e. the cumulative square error (CSE)
\begin{eqnarray}
{CSE} = \sum_{n=0}^N \Big(p(u|v')-p(u|v^{'(n)})\Big)^2.
\end{eqnarray}
The objective is to minimize this cumulative square error and so we will compute and store the cumulative error for evaluation purposes during testing.

We will test a few algorithms exhaustively over all possible $2^8$ states of the Huang network and exhaustively over all possible nontrivial (joint) probabilistic inquiries for testing purposes (all set combinations of $U$ and $V$). Because some $v'$ are more likely than others, we do a weighted average of the cumulative square error and take a square root to generate a final score, which we call the cumulative root mean squared error (CRMSE), 
\begin{eqnarray}
{CRMSE} = \sum_{n=0}^N\sqrt{\sum_{v'\in V}p(v')\Big(p(u|v')-p(u|v^{'(n)})\Big)^2},\label{CRMSE}
\end{eqnarray}
which is reported in Tables \ref{table2} and \ref{table3} along with the $n$th RMSEs, $$\sqrt{\sum_{v'\in V}p(v')(p(u|v')-p(u|v^{'(n)}))^2}.$$

\subsubsection{Algorithms}

We implemented 1-node and 2-node look ahead greedy algorithms using the MI and the partial inferential deviation (PID) (\ref{partial_inequality}) and compared results in terms of (\ref{CRMSE}). Starting from a (joint) marginal probability of interest $p(u)$, the 1-node look ahead greedy PID (MI) algorithm chooses which node from $V$ for the sensor to measure next based on the evaluation of the PID (MI). The selected node is measured, which updates the distribution to a posterior, which becomes the new estimate of the state probability of interest $p(u|v^{'(n)})$. The sensor selection process starts over given the updated state, which results in the previously mentioned ``inference trajectories". The 2-node look ahead greedy algorithms, 2-PID and 2-MI, select which pairs of nodes (given there were 2 or more remaining taskable nodes) are expected to remove to the most PID or MI and then select which node from the selected pair has a higher single node PID or MI. Both algorithms reevaluate their next measurement choice given the measured outcomes from previous choices.

The maximum MI algorithms use the $u$th term of the MI, ${MI}_{u}$, from the sum, $${MI}=\sum_{u\in U}{MI}_{u} \equiv \sum_{u\in U}p(u)\Big(\sum_{v\in V}p(v|u)\ln(p(v|u)/p(v))\Big)=\sum_{u\in U}p(u){IG}[u=U,V],$$ 
as we are evaluating how much $V$ informs $u$. Maximizing ${MI}_{u}$ with respect to the choice of various subsets of $V$ is, without loss of optimality, equal to optimizing the information gain ${IG}[u=U,V]$ as the $p(u)$ factors out. Mutual informations are evaluated at each stage in the inference trajectory by considering the informativeness of each possible available measurement (or two possible measurements in the 2-node look ahead case), conditional on what has already been observed $v^{'(n)}$.

Similarly, the maximum PID algorithms evaluate PIDs at each stage in the inference trajectory by considering the amount of inferential deviation each possible available measurement (or two possible measurements in the 2-node look ahead case) is expected to contribute, conditional on what has already been observed $v^{'(n)}$. Due to the relation to the MSE in equation (\ref{inferential_variance}), the PID algorithms are expected to improve over the MI in terms of CRMSE, which we will show to be generally true in the results. However, because the optimization problem is NP-hard \cite{vanslette2023conjunction}, the greedy algorithms we implemented only approximate the true optimal sensor tasking strategy. 
\subsubsection{Results}

\begin{table}
\begin{center}
\begin{tabular}{ll|llllllll}
\hline
\multicolumn{2}{c|}{Case } & \multicolumn{7}{c}{Posterior probability ${RMSE}$ after $n\leq N$ measurements (conditions) } \\
\hline
Dim. & Alg. & CRMSE & $n=0$ & $n=1$ & $n$ = 2 &$n$ =3 &$n$ = 4 & $n$ = 5 \\
\hline
1 & PID& 0.7597 & 0.3386 & \textbf{0.2101} & 0.1140& 0.0719 & 0.0250 & 0.0 \\
& \textbf{MI}& \textbf{0.7499} & & 0.2116 & \textbf{0.1111} & \textbf{0.0638} & \textbf{0.0249} & 0.0 \\
\hline
2 & \textbf{PID}& \textbf{0.6327} & 0.2607 & \textbf{0.1813} & \textbf{0.1120} & \textbf{0.0532} & \textbf{0.0204}& \textbf{0.0052} \\
& MI& 0.6384 & & 0.1820 & 0.1131 & 0.0545 & 0.0225& 0.0056 \\
\hline
3 & \textbf{PID} & \textbf{0.4242} & 0.1712 & \textbf{0.1230} & \textbf{0.0800} & \textbf{0.0378} & \textbf{0.0122} & \\
& MI &0.4267 & & 0.1234 & 0.0814 & 0.0382 & 0.0125 & \\
\hline
4 &\textbf{PID}& \textbf{0.2395} & 0.1034& \textbf{0.0734} & \textbf{0.0454}& 0.0174 & & \\
& MI & 0.2399 & & 0.0737 & 0.0455 & \textbf{0.0172} & & \\
\hline
5 & \textbf{PID}& \textbf{0.1164} & 0.0580 & \textbf{0.0385} & 0.0199 & & & \\
& MI & 0.1165 & & 0.0387 & \textbf{0.0197} & & & \\
\hline
6 & \textbf{PID}& \textbf{0.0461} & 0.0297& \textbf{0.0164} & & & & \\
& MI&0.0462 & & 0.0165 & & & & \\
\hline
\end{tabular}

\caption{\label{table2} A posterior probability to ground truth probability ${RSME}$ comparison of the greedy maximum partial inferential deviation (${PID}$) algorithm against a greedy maximum mutual information algorithm (${MI}$) for a number of cases. Bolded is the algorithm that has superior performance for a case. The ``Dim." column indicates the dimension (i.e. number of nodes) of $U$ we are making joint inferences about in the given case while $n$ represents the number of sensors we tasked that made measurements. For example $p(f,h|a')$ is 2 dimensional with $n=1$ measured values. The $n=0$ case is the starting ${RMSE}$ between the marginal distribution (prior) and the actual conditional distribution before any sensors are tasked to make measurements. The ${RSME}$ quantifies the average (root mean square) error of the probability estimates. CRMSE's are computed prior to rounding. }

\end{center}
\end{table}

The 1-node look ahead greedy algorithms are compared in Table \ref{table2}. The PID algorithm improves upon in MI in terms of reducing the inferential error of posterior predictions in most of the test cases shown here; however, the two algorithms often gave equal inference trajectories for this relatively simple case. Upon inspection, the examples in the 1 dimensional case that the PID algorithm failed to improve upon were a subset of the trajectories corresponding to the state $p(e)$. Due to normalization, the inferential variation of a binary variable is equal for both states ${Var}_B\Big[p(a=0|b)\Big] ={Var}_B\Big[1-p(a=1|b)\Big] ={Var}_B\Big[p(a=1|b)\Big] $, which means this greedy PID algorithm will choose the same nodes independent of whether the state in question is 0 or 1 while the MI algorithm (max ${MI}_{a}$) is not restricted in this way. Once the PID algorithm starts dealing with joint states (case 2 and above) it is no longer overly constrained by normalization and it outperforms the MI algorithm in terms of ${RSME}$ nearly everywhere.

\begin{table}
\begin{center}
\begin{tabular}{ll|llllllll}
\hline
\multicolumn{2}{c|}{Case } & \multicolumn{7}{c}{Posterior probability ${RMSE}$ after $n\leq N$ measurements (conditions) } \\
\hline
Dim. & Alg. & CRMSE & $n=0$ & $n=1$ & $n$ = 2 &$n$ =3 &$n$ = 4 & $n$ = 5 \\
\hline
\hline
1&\textbf{2-PID} & \textbf{0.7388} & 0.3386 & 0.2130 & \textbf{0.1081} & \textbf{0.0545} & 0.0247&0.0 & \\
&2-MI & 0.7498 & & \textbf{0.2116} & 0.1111 & 0.0638 & 0.0247 &0.0 & \\
\hline
2&\textbf{2-PID} &\textbf{0.6310} & 0.2607 & \textbf{0.1825} & 0.1105 & \textbf{0.0530}& \textbf{0.0196} & 0.0047 & \\
& 2-MI &0.6320 & & 0.1839 & \textbf{0.1093} & 0.0537 & 0.0206 & \textbf{0.0039} & \\
\hline
3&\textbf{2-PID} & \textbf{0.4213} & 0.1712 & \textbf{0.1238} & \textbf{0.0788} & \textbf{0.0362} & \textbf{0.0113} && \\
& 2-MI & 0.4220 & & 0.1240 & 0.0791 & 0.0363 & 0.0114 & & \\
\hline
4&\textbf{2-PID} & \textbf{0.2382} & 0.1034 & 0.0740 & \textbf{0.0448} & 0.0161& & & \\
& 2-MI & 0.2387 & & 0.0740 & 0.0452 & 0.0161 & & & \\
\hline
5&\textbf{2-PID} &\textbf{0.1161} & 0.0580 & \textbf{0.0387} & \textbf{0.0194} & & & & \\
& 2-MI & 0.1163 & & 0.0388 & 0.0195& & & & \\
\hline
6&\textbf{2-PID} & \textbf{0.0461} & 0.0297& \textbf{0.0164} & & & & \\
& 2-MI&0.0462 & & 0.0165 & & & & & \\
\hline

\end{tabular}

\caption{\label{table3} Results comparing the 2-PID and 2-MI sensor tasking algorithms in terms of posterior state probability estimation RMSE. Because the two-step greedy algorithm reduces to the one-step greedy algorithm in the last row, the results in the last row of the tables are equal. }

\end{center}
\end{table}

The 2-node look ahead greedy algorithms are compared in Table \ref{table3}. The 2-PID algorithm out performs the 2-MI algorithm in cumulation, although the 2-MI algorithm occasionally beats the 2-PID algorithm on a case-by-case basis for some evaluations. Further, as we expect, the 2-PID algorithm outperforms the PID algorithm in all cases for $n>1$. The 2-MI algorithm also performed better than the MI algorithm in terms of RMSE; however, 2-PID improved more over the PID than the 2-MI improved over the MI. We attribute this to the 2-PID algorithm better approximating the optimal tasking solution in terms of RMSE while the 2-MI instead is better approximating the optimal MI solution. Thus, we attribute the MI algorithm's success over the PID in row one of Table \ref{table2} to be partially attributable to an ``accumulation of circumstance" coming from the NP-hard nature of the optimization problem along with the inferential deviation being constrained by normalization as explained earlier. Both of these point to a need for more rigorously algorithms for inferential RMSE-based sensor tasking; however, these results can be used as a baseline in future studies.

\section{Conclusions}

The research presented uses Bayesian inference to make statements about its own behavior to improve probabilistic reasoning and expand the Bayesian inference framework. We treated Bayesian probability updating as an uncertain probability updating process prior to measurement. Through the quantification of inferential moments and the inferential probability distribution, we were able to quantify and describe new intrinsic properties of joint probability distributions. 

In particular, we demonstrated that inferential deviations are a key tool for understanding the expected variation of the probability of one variable in response to an inferential update of another. This means that the inferential deviation can be used to expresses one's uncertainty or mean squared error in a probabilistic estimate due to unknown but definite values of another variable, $p(a)\pm\sigma_B\Big[p(a|b)\Big]$. Because probabilistic graphical models like Bayesian networks define a joint probability distribution, one can compute inferential moment information to enhance situational awareness (e.g. by quantifying its inferential deviations) as demonstrated in our first application. In our second application, we compared greedy inferential deviation based algorithms for optimal sensor tasking to mutual information based algorithms. Our algorithm generally outperformed the mutual information algorithm in terms of probabilistic ${RMSE}$, which we argue is more useful for state estimation and sensor tasking than is amounts of information. The inferential deviation approach here offers an error reducing inference approach to compete with informational-based inference approaches for these types of tasks. This is analogous to how least squares and maximum likelihood offer competing solutions for regression and classification problems by optimizing different cost functions. The sensor tasking algorithms we presented are rudimentary and likely can be improved with more sophisticated techniques as well as incorporate either higher order inferential moments or inferences from the inferential probability distribution. 

This extended inference framework provides new tools to reason about the properties and behavior of probability distributions themselves similar in function to information theoretic quantities. Supporting this claim is the existence of a power series expansion of the mutual information in terms of inferential moments, the relationship of the inferential variance to the mean squared error, and the results and ability to formulate sensor tasking applications purely within a Bayesian inference framework. There are several avenues for future research using this approach including the creation algorithms for feature selection and Bayesian network structure learning, improved mathematic rigor and study, and continuing the comparison between Bayesian inferential theoretic and information theoretic approaches.



%

\vspace{6pt}

\paragraph{Funding}
This research received no external funding.

\paragraph{Acknowledgments}
Carl Andersen, Gerasimos Angelatos, Nick Carrara, Ariel Caticha, Zac Dutton, Tony Falcone, Jeremie Fish, Roddy Taing, Tony Tohme, and Yuri Strohm



\paragraph{Abbreviations}
The following abbreviations are used in this manuscript:\\

\noindent 
\begin{tabular}{@{}ll}
2-MI & Mutual information two-step look-ahead greedy algorithm\\
2-PID & Partial inferential deviation two-step look-ahead greedy algorithm\\
Alg. & Algorithm\\
AI & Artificial intelligence\\
CSE & Cumulative squared error\\
CRMSE & Cumulative root mean squared error\\
Dim. & Dimension of state variables in question\\
IG & Information gain\\
MI & Mutual information (value or one step look ahead algorithm depending on the context)\\
MSE & Mean squared error\\
PID & Partial inferential deviation one step look ahead algorithm\\
RMSE & Root mean squared error
\end{tabular}

\appendix
\section[\appendixname~\thesection]{Appendix}\label{appendix1}
Due to (\ref{expected_marginal}), the marginal posterior Bayes' rule (\ref{marginalPosteriorBayes}) can instead be written as a Bayes' rule update of the inferential expectation. That is, $p(a)\stackrel{*}{\rightarrow}p'(a)$ is equally written as,
\begin{eqnarray}
{E}_{B}\Big[p(a|b)\Big]\stackrel{*}{\rightarrow}{E}_{B}\Big[p(a|b)\Big|b'\Big],
\end{eqnarray}
where the posterior inferential expectation is,
\begin{eqnarray}
{E}_{B}\Big[p(a|b)\Big|b'\Big]=\sum_{b} p(a|b) p(b|b') =\sum_{b} p(a|b) \delta_{b,b'}=p(a|b')= p'(a).
\end{eqnarray}
We used the standard notation for conditional expectation, as given in the notation subsection of the background, $${E}_{X}\Big[f(x)\Big|y\Big] \equiv \sum_{x\in X}f(x)p(x|y),$$ where here $X=B$, $f(x)=p(a|b)$, $y=b'$, and $p(x|y)=p(b|b')$.

Given this, Bayes' Rule can be used analogously to update higher order inferential moments in a straightforward manner, i.e.,
\begin{eqnarray}
{E}_{B}\Big[p(a|b)^n\Big]\stackrel{*}{\rightarrow}{E}_{B}\Big[p(a|b)^n\Big|b'\Big]=p(a|b')^n,\label{bayesn}
\end{eqnarray}
and similarly for central inferential moments,
\begin{eqnarray}
{E}_{B}\Big[(p(a|b)-p(a))^n\Big] \stackrel{*}{\rightarrow} {E}_{B}\Bigg[\Big(p(a|b)-{E}_{B}\Big[p(a|b)\Big|b'\Big]\Big)^n\Bigg|b'\Bigg]=(p(a|b')-p'(a))^n=0.
\end{eqnarray}
The inferential probability distribution can be inferentially updated, and for a single variable in question yields a singular distribution for $p'(a)$,
\begin{eqnarray}
\mathcal{P}_B[p'(a)]\stackrel{*}{\rightarrow}\mathcal{P}_B[p'(a)|b'] = \delta_{p'(a),p(a|b')}\,,
\end{eqnarray}
which similarly proves that all of the posterior inferential central moments of a single updating variable are zero (i.e. due to (\ref{MGF})), which is reasonable as there is no longer any inferential uncertainty from $B$ as $B$ was measured. Things are less trivial in higher dimensions. 

Consider the inference of $a\in A$ given two uncertain measurable variables $b\in B$ and $c\in C$ and the conditional distribution $p(a|b,c)$. The inferential expectation, in response to a measurement from $B$, updates as follows:\footnote{Technically $p(b,c)\stackrel{*}{\rightarrow}p(b,c|b') = p(c|b,b')p(b|b')= p(c|b')p(b|b',c)$. However, because $B'$ reprents a direct and completely certain measured value corresponding to the ``real" underlying variable value $B$, the probability $p(b|b',c)$ is naturally conditionally independent of $c$ given $b'$, i.e. $p(b|b',c)=\delta_{b,b'}=p(b|b')$. Consequently, this implies $p(c|b,b')=p(c|b')$, which models $p(c|b')=p(c|b)$ for $b'=b$. }
\begin{eqnarray}
{E}_{B,C}\Big[p(a|b,c)\Big]\stackrel{B}{\rightarrow}{E}_{B,C}\Big[p(a|b,c)\Big|b'\Big] = \sum_{b,c}p(a|b,c)p(c|b')\delta_{b,b'} = p(a|b').\label{reduceupdaterule0}
\end{eqnarray}
The rule for inferentially updating an $n$th order central inferential moment from prior to posterior using Bayes' Rule follows similarly:
\begin{eqnarray}
{E}_{B,C}\Big[(p(a|b,c)-p(a))^n\Big] \stackrel{B}{\rightarrow} {E}_{B,C}\Bigg[\Big(p(a|b,c)-{E}_{B,C}\Big[p(a|b,c)\Big|b'\Big]\Big)^n\Bigg|b'\Bigg],
\end{eqnarray}
which reduces to:
\begin{eqnarray}
{E}_{B,C}\Big[(p(a|b,c)-p(a))^n\Big] \stackrel{B}{\rightarrow} {E}_{C}\Big[\Big(p(a|b',c)-p(a|b')\Big)^n\Big],\label{reduceupdaterule}
\end{eqnarray}
which is nothing but a way of writing central inferential moments over an updated posterior distribution. The proof of equation (\ref{reduceupdaterule}) is,
\begin{eqnarray}
{E}_{B,C}\Bigg[\Big(p(a|b,c)-{E}_{B,C}\Big[p(a|b,c)\Big|b'\Big]\Big)^n\Bigg|b'\Bigg]&=&{E}_{B,C}\Big[(p(a|b,c)-p(a|b'))^n\Big|b'\Big]\nonumber\\
&=&\sum_{b,c}(p(a|b,c)-p(a|b'))^np(c|b')\delta_{b,b'}\nonumber\\
&=&\sum_{c}(p(a|b',c)-p(a|b'))^np(c|b')\nonumber\\
&=&{E}_{C}\Big[\Big(p(a|b',c)-p(a|b')\Big)^n\Big|b'\Big].\label{reduceupdaterule2}
\end{eqnarray}
Note the posterior inferential variance is,
\begin{eqnarray}
{E}_{C}\Big[\Big(p(a|b',c)-p(a|b')\Big)^2\Big|b'\Big]&=&\sum_{c}(p(a|b',c)-p(a|b'))^2p(c|b')\nonumber\\
&=& {Var}_{C}\Big[p(a|b',c)\Big|b'\Big],\label{posteriorinfvar}
\end{eqnarray}
as the measure, $p(c|b')$, and the argument, $p(a|b',c)$, are both conditional on $b'$. Further,
The inferential probability distribution in the two variable case is,
\begin{eqnarray}
\mathcal{P}_{B,C}[p'(a)] = \sum_{b\in B} \sum_{c\in C} \delta_{p'(a),p(a|b,c)}\, p(b,c). 
\end{eqnarray}
For an update of just $B$, the inferential probability distribution updates as follows:
\begin{eqnarray}
\mathcal{P}_{B,C}[p'(a)]\stackrel{B}{\rightarrow}\mathcal{P}_{B,C}[p'(a)|b'] =\sum_{c\in C} \delta_{p'(a),p(a|b',c)}\, p(c|b')
\end{eqnarray}
which is the inferential probability distribution of $a$ due to $c$ conditioned on $b'$.

Let's now derive the inferential variance decomposition formula (\ref{total_inferential_variance}). Consider the difference in inferential variance between the 2 dimensional and 1 dimensional cases:
\begin{eqnarray}
\bigtriangleup \equiv {Var}_{B,C}\Big[p(a|b,c)\Big]-{Var}_{B}\Big[p(a|b)\Big].
\end{eqnarray}
Expand both expressions in terms of probabilities using ${Var}[x] = {E}[x^2] - {E}[x]^2$
\begin{eqnarray}
\bigtriangleup &=&\Big[\sum_{b,c} p(a|b,c)^2p(b,c) - p(a)^2\Big]-\Big[ \sum_b p(a|b)^2p(b) - p(a)^2\Big]\nonumber\\
&=&\sum_{b,c} p(a|b,c)^2p(b,c) - \sum_b p(a|b)^2p(b)\nonumber\\
&=&\sum_{b,c} p(a|b,c)^2p(b,c) - \sum_{b,c} p(a|b)^2p(b,c)\label{lasteq}\\
&=&\sum_{b,c} \Big(p(a|b,c)^2 - p(a|b)^2\Big) p(b,c)\nonumber\\
&=&\sum_bp(b)\sum_{c} \Big(p(a|b,c)^2 - p(a|b)^2\Big) p(c|b)\nonumber\\
&=&{E}_B\Big[\sum_{c} \Big(p(a|b,c)^2 - p(a|b)^2\Big) p(c|b)\Big]\nonumber\\
&=&{E}_B\Big[\sum_{c} \Big(p(a|b,c)^2p(c|b)\Big) - {E}_C\Big[p(a|b,c)\Big|b\Big]^2\Big] \label{lasteq23}\\
&=&{E}_B\Bigg[{Var}_{C}\Big[p(a|b,c)\Big|b\Big]\Bigg]\geq 0.\label{last3}
\end{eqnarray}
Equation (\ref{lasteq}) uses $p(b)=\sum_c p(b,c)$, equation (\ref{lasteq23}) uses, $$p(a|b) = \sum_{c}p(a,c|b)= \sum_{c}p(a|b,c)p(c|b)= {E}_C\Big[p(a|b,c)\Big|b\Big],$$ and equation (\ref{last3}) is the conditional variance identity. Note the final expression is the the posterior inferential variance, equation (\ref{posteriorinfvar}). This gives what we will call the inferential variance decomposition formula, 
\begin{eqnarray}
{Var}_{B,C}\Big[p(a|b,c)\Big]={E}_B\Bigg[{Var}_{C}\Big[p(a|b,c)\Big|b\Big]\Bigg]+{Var}_{B}\Big[p(a|b)\Big].\label{inferential variance decomposition formula}
\end{eqnarray}

The inferential variance decomposition formula can also be derived from the law of total variance, but in a bit of a round-about way. The law of total variance is,
\begin{eqnarray}
{Var}_{Y}[y] = {E}_X\Big[{Var}_{Y}[y|x]\Big] + {Var}_{X}\Big[{E}_Y[y|x]\Big],
\end{eqnarray}
which also holds for functions of $y\rightarrow f(y)$,
\begin{eqnarray}
{Var}_{Y}[f(y)] = {E}_X\Big[{Var}_{Y}[f(y)|x]\Big] + {Var}_{X}\Big[{E}_Y[f(y)|x]\Big].
\end{eqnarray}
Letting $f(y)\rightarrow p(a|b,c)$, $Y\rightarrow B\times C$ and $X\rightarrow B'$ be a perfect measurement variable of $B$ such that $p(b)=p(b')$ for $b=b'$ and $p(b|b')=\delta_{b,b'}$, the law of total variance is 
\begin{eqnarray}
{Var}_{B,C}\Big[p(a|b,c)\Big] &=& {E}_{B'}\Bigg[{Var}_{B,C}\Big[p(a|b,c)\Big|b'\Big]\Bigg] + {Var}_{B'}\Bigg[{E}_{B,C}\Big[p(a|b,c)\Big|b'\Big]\Bigg].
\end{eqnarray}
The following expressions can be reduced using equation (\ref{reduceupdaterule0}), $${E}_{B,C}\Big[p(a|b,c)\Big|b'\Big]= p(a|b'),$$ and equations (\ref{reduceupdaterule2},\ref{posteriorinfvar}), $${Var}_{B,C}\Big[p(a|b,c)\Big|b'\Big]={Var}_{C}\Big[p(a|b',c)\Big|b'\Big].$$ Because $B'$ is distributed the same as $B$, this shows that the inferential variance decomposition formula is an odd form of the law of total variance as making these substitutions yields (\ref{inferential variance decomposition formula}),
\begin{eqnarray}
{Var}_{B,C}\Big[p(a|b,c)\Big]&=&{E}_{B'}\Bigg[{Var}_{C}\Big[p(a|b',c)\Big|b'\Big]\Bigg]+{Var}_{B'}\Big[p(a|b')\Big]\nonumber\\
&=&{E}_B\Bigg[{Var}_{C}\Big[p(a|b,c)\Big|b\Big]\Bigg]+{Var}_{B}\Big[p(a|b)\Big].
\end{eqnarray}

\bibliographystyle{unsrt}  
\bibliography{templatePRIME} 

\end{document}